\documentclass[reprint,superscriptaddress
]{revtex4-2}
\usepackage{amsmath}
\usepackage{amssymb}
\usepackage[dvipdfmx]{graphicx}
\usepackage{bm}
\usepackage{mathtools}
\newcommand{\B}[1]{{\bm #1}}%数式中のボールド
\newcommand{\mac}[1]{{\mathcal #1}}%Hamiltonianとか
\newcommand{\mab}[1]{{\mathbb #1}}%Hamiltonianとか
\newcommand{\eref}[1]{Eq.\ (\ref{#1})}%式の参照
\newcommand{\dref}[1]{Fig.\ \ref{#1}}%図の参照
\newcommand{\T}{\mathrm{T}}%転置

\newcommand{\btheta}{\bm \theta}
\newcommand{\pdif}[2]{\dfrac{\partial #1}{\partial #2}}%上の微分の偏微分バージョン
\newcommand{\nn}{\nonumber}%式番号を振らない

\begin{document}

\title{Statistical-mechanical study of deep Boltzmann machine given weight parameters after training by singular value decomposition}

\author{Yuma Ichikawa$^1$\thanks{ichikawa-yuma1@g.ecc.u-tokyo.ac.jp} and Koji Hukushima$^{1, 2}$\thanks{k-hukushima@g.ecc.u-tokyo.ac.jp}}
\affiliation{$^1$ Graduate School of Arts and Sciences, The University of Tokyo, Meguro, Tokyo 153-8902, Japan \\
$^2$ Komaba Institute for Science, The University of Tokyo, Meguro, Tokyo 153-8902, Japan}

\begin{abstract}
Deep learning methods relying on multi-layered networks have been actively studied in a wide range of fields in recent years, and deep Boltzmann machines(DBMs) is one of them.
In this study, a model of DBMs with some properites of weight parameters obtained by learning is studied theoretically by a statistical-mechanical approach based on the replica method.
The phases characterizing the role of DBMs as a generator and their phase diagram are derived, depending meaningfully on the numbers of layers and  units in each layer. It is found, in particular, that the correlation between the weight parameters in the hidden layers plays an essential role and that an increase in the layer number has a negative effect on DBM's performance as a generator when the correlation is smaller than a certain threshold value.
\end{abstract}

\maketitle

\section{Introduction}
In recent years, deep-learning-based  methods have been studied in a wide range of fields. One such implementation of them is in  deep Boltzmann machines(DBMs)\cite{salakhutdinov2009deep}.
DBM is a deep undirected graphical model,
%an extension of deep belief network\cite{hinton2006fast}, a probabilistic model defined on a layered netwrok, to an undirected graph,
which has a layered structure of restricted Boltzmann machines (RBMs)\cite{smolensky1986information, hinton2002training}, i.e., Boltzmann machines(BMs)\cite{ackley1985learning,wainwright2008graphical,neubig2012information} with simplified topology.
Because DBMs allow extracting the hierarchical features of given data and representing complex probability distributions, they have been effectively used in a variety of fields such as generating artificial data\cite{eslami2014shape}, multimodal learning\cite{srivastava2012multimodal}, image recognition\cite{leng20153d}, and pretraining for deep feedforward networks (DNNs)\cite{erhan2010does}. Furthermore, DBMs are considered to have  more powerful variational representations of many-body wavefunctions\cite{gao2017efficient} than RBMs\cite{melko2019restricted, carleo2017solving}.

Despite these practical applications, DBMs still have some limitations, causing them to be overshadowed by DNNs.
The exact learning of DBMs is computationally difficult because they include evaluations of intractable summations, and thus empirical approximate learning methods have been proposed\cite{salakhutdinov2009deep,salakhutdinov2010efficient,hinton2012better, cho2013gaussian}. However, the theoretical validity of such approximate learning methods remains unclear.
In addition, these algorithms require high speed and accuracy.
Furthermore, compared with RBMs, the learning of DBMs is markedly affected by the problem of local optimum solutions. To avoid this, various methods for finding appropriate initial values of the parameters - called pretraining -  %is necessary. Various pretraining methods
have been proposed\cite{salakhutdinov2010efficient,hinton2012better,cho2013two}, but the computational complexity increases significantly as the number of layers increases. Therefore, it is important to propose a more accurate and faster pretraining method.
Moreover, it is difficult to determine the optimal network structure to obtain sufficient accuracy for a given task, and the methodology has not been sufficiently discussed.

To address these issues, it is important to clarify the required DBM settings for DBMs to generate samples that are similar to the given data. In the settings, clarifying the effect of constructing a layered network on learning and representation abilities is significant. Also, the configurations of DBMs, in particular the number of layers and the number of elements in each layer, are important for practical applications. Such settings are generally considered to depend on given data, but it is expected that there are also typical properties independent of the data.
In this study, we focused on these typical properties from the viewpoint of statistical mechanics of random systems, which may provide new implications for the learning and pretraining methods of DBM.

Because of the clear correspondence between BMs and statistical mechanics of random systems, there have been analyses of the typical properties of RBMs using methods of the statistical mechanics of random systems\cite{marullo2020boltzmann, decelle2021restricted}.
In order to evaluate the typical properties, it is necessary to introduce an ensemble into RBMs, and previous studies can be mainly categorized into two types of ensembles.
One is that the weight parameters of RBMs itself follow a random ensemble, and the other is that the weight parameters are represented by a singular-value decomposition  and their singular vectors follow a random ensemble.
The former analyses revealed that RBMs can learn higher-order correlations\cite{tubiana2017emergence, tubiana2019learning, agliari2012multitasking, barra2017phase,barra2018phase, mezard2017mean, agliari2020tolerance} and that an RBM generating random patterns with multiple features corresponds to a Hopfield model storing multiple patterns in parallel% and a duality relation between Hopfield model and RBMs have been proved
\cite{barra2012equivalence, marullo2020boltzmann}.
The latter study can be considered as RBMs under more realistic settings, and its properties as a generator are discussed in relation to a thermodynamic phase diagram\cite{decelle_spectral, decelle2018thermodynamics}.

As for DBMs, there are some studies that have considered an ensemble of random weight parameters and discussed the mathematical properties of DBMs without considering the complex correlations among the elements of the weight parameters by learning
\cite{genovese2020minimax, alberici2020annealing, alberici2021solution, alberici2021deep, agliari2021pattern}.
However, the assumption of weight parameters whose elements are independent and identically distributed is somewhat unrealistic because it does not take into account the correlations of the weight parameters that result from learning.
Meanwhile, there has been no analysis of DBMs in which the weight parameters are represented by singular-value decomposition as discussed in RBMs.

In this study, we conducted numerical experiments to investigate the properties of DBMs to function well as a generator, and we used a statistical mechanical method to analyze the DBM, in which the weight parameters are expressed by singular-value decomposition based on the results of numerical experiments.
Our analysis reveals that % and we also clarified
the properties of the weight parameters for DBMs to typically function well as generators, as well as the effect of the layered network structure of DBMs, that is absent in RBMs.

This paper is structured as follows.
In Sect.~\ref{sec:learning}, the DBM and its training method are described.
In Sect.~\ref{sec:sec3}, we introduce the representation by singular-value decomposition of the weight parameters after learning and explain our numerical experiments that bridge the modeling of DBM to analyze the typical properties.
In Sect.~\ref{sec:model}, we present a model based on the numerical experiments presented in the previous section.
In Sect.~\ref{sec:replica}, the model is analyzed using the replica method, which is a statistical mechanical method for random systems.
In Sect.~\ref{sec:PD}, we derive the phase diagram of a 3-layer DBM, which is the simplest model with a hierarchical structure, and clarify the conditions for the 3-layer DBM to function well as a generator.
In particular, we discuss the dependence of the phase diagram on the number of units in each layer and the correlations between the weight parameters of each layer that appear because of the coupling between the hidden layers.
In Sect.~\ref{sec:infinite}, we discuss the dependence of the phase diagram on the number of hidden layers and discuss the convergence of the phase boundary of DBM in the limit of infinite layer number.
Sect.~\ref{sec:summary} is devoted to the conclusions of this paper and future prospects.

\section{DBM and its Learning Procedure}
\label{sec:learning}
DBMs discussed in this paper consist of a visible layer and $L$ hidden layers. The visible layer has $N_{0}$ visible units
$\B{\sigma}^{(0)} = (\sigma_{i}^{(0)} )_{i=1}^{N_{0}}$ and $l$-th
hidden layer has $N_{l}$ hidden units
$\B{\sigma}^{(l)} =(\sigma^{(l)}_{j} )_{j=1}^{N_{l}}$. All units are Ising variables $\sigma_{i}^{(l)} \in \{-1, 1\}$ and the set of all units is denoted by $\B{\sigma} = \{\B{\sigma}^{(l)}\}_{l=0}^{L}$. The energy function is defined as follows:
\begin{equation}
  \label{eq:energy}
  E(\bm{\sigma} ; \bm{\theta}) \coloneqq - \sum_{l=1}^{L} (\B{\sigma}^{l-1})^{\top} \B{W}^{(l)} \B{\sigma}^{(l)},
\end{equation}
where $\B{W}^{(l)} = (W_{ij}^{(l)})$ is a weight parameter for $\B{\sigma}^{(l-1)}$ and $\B{\sigma}^{(l)}$ and $\top$ denotes the transposition of the matrix. The set of weight parameters is denoted by $\B{\theta}$.

The joint distribution of all the units in DBM is defined by the energy function in \eref{eq:energy} as
\begin{equation}
\label{eq:dbm-canonical}
    p(\B{\sigma} \mid \btheta) \coloneqq \frac{1}{Z(\btheta, \beta)} e^{- \beta E(\B{\sigma} ; \btheta)},
\end{equation}
where the inverse temperature parameter $\beta$ is a positive real number
and the normalization constant $Z(\btheta, \beta)$ is defined by
\begin{equation}
Z(\btheta, \beta) \coloneqq \sum_{\B{\sigma}} e^{ - \beta E(\B{\sigma} ; \btheta)},
\end{equation}
where $\sum_{\B{\sigma}}$ is the sum over all possible realizations of $\B{\sigma}$. It is possible to add linear fields,
called bias terms, to each unit, but we considered the case without bias terms in this study. When $L=1$, DBM is equivalent to RBM without bias terms.

The training of DBMs involves determining the weight parameters $\B{\theta}$ for a given data $\mac{D} = \{\B{\sigma}^{\mu, (0)}\}_{\mu=1}^{P}$, which is generally performed by maximizing
the log-likelihood of DBMs expressed by
\begin{equation}
  \label{eq:log-likelihood}
  \mac{L}_{\mac{D}}(\B{\theta}) \coloneqq \frac{1}{P} \sum_{\mu=1}^{P} \ln \sum_{\B{\sigma} \setminus \B{\sigma}^{(0)}} p(\B{\sigma}^{\mu, (0)}, \B{\sigma}\setminus\B{\sigma}^{(0)} \mid \B{\theta}).
\end{equation}
The gradients of the log-likelihood with respect to the weight parameters for $l = 1, \ldots, L$ are given by
\begin{multline}
  \label{eq:gradient-log-likelihood}
  \frac{\partial \mac{L}_{\mac{D}}(\B{\theta})}{\partial W_{ij}^{(l)}} = \frac{1}{P} \sum_{\mu=1}^{P} \sum_{\B{\sigma} \setminus \B{\sigma}^{(0)}} h_{i}^{(l-1)} h_{j}^{(l)} p(\B{\sigma} \setminus \B{\sigma}^{(0)} \mid \B{\sigma}^{\mu, (0)}, \B{\theta}) \\
  - \sum_{\B{\sigma}} h_{i}^{(l-1)} h_{j}^{(l)} p(\B{\sigma} \mid \B{\theta}).
\end{multline}
Maximizing the \eref{eq:log-likelihood} is computationally difficult because the gradients in \eref{eq:gradient-log-likelihood} involve intractable summations over all units. Hence, maximization is performed using approximate algorithms. In conventional methods\cite{salakhutdinov2009deep,salakhutdinov2010efficient}, the first term of \eref{eq:gradient-log-likelihood} is evaluated by mean-field approximation, whereas the second term of \eref{eq:gradient-log-likelihood} is evaluated by Gibbs sampling and persistent contrastive divergence(PCD) proposed in Ref.~\cite{tieleman2008training}. However, compared with RBMs, the problem of local optimal solutions is more serious in learning DBMs. Therefore, it is essential to use a method to search for the appropriate initial values of the weight parameters, which is generally called pretraining.
Various methods have been proposed for the pretraining of DBMs and greedy learning by RBMs trained by the contrastive divergence(CD) method\cite{hinton2002training} or the PCD method is widely used\cite{hinton2012better, salakhutdinov2010efficient}.

\section{Numerical Experiments}
\label{sec:sec3}
In the previous statistical-mechanical studies of RBMs\cite{decelle2017spectral, decelle2018thermodynamics},
numerical experiments suggested that
the first few singular modes of the weight parameters were significant for RBMs to function well as generators, and
the remaining modes behaved as noise.
In this study, we also conducted numerical experiments
to learn the weight parameters of DBMs and perform singular-value decomposition of the learned weight parameters $\B{W}^{(l)}$ of DBMs,
defined by
\begin{equation}
  \B{W}^{(l)} = \B{U}^{(l)} \B{\Sigma}^{(l)} (\B{V}^{(l)})^{\T},
\end{equation}
where $\B{U}^{(l)}$ is an $N_{l-1} \times N_{l}$ orthogonal matrix consisting of the left singular vectors $\B{u}^{(l)}_{\alpha_{l}},~(\alpha_{l} = 1, \ldots, N_{l})$ in each column, $\B{V}^{(l)}$ is an $N_{l}\times N_{l}$ orthogonal matrix consisting of right singular vectors $\B{v}^{(l)}_{\alpha_{l}}$ in each column, and $\B{\Sigma}^{(l)}$ denotes a diagonal matrix with diagonal elements of singular values $w^{(l)}_{\alpha_{l}}$
in descending order of $w_{1} \ge w_{2} \ge \cdots$. Each singular vector satisfies the following equations:
\begin{align}
  \B{W}^{(l)} \B{v}^{(l)}_{\alpha_{l}} &= w^{(l)}_{\alpha_{l}} \B{u}^{(l)}_{ \alpha_{l}},\\
  (\B{W}^{(l)})^\T \B{u}^{(l)}_{\alpha_{l}} &= w^{(l)}_{\alpha_{l}} \B{v}^{(l)}_{ \alpha_{l}}.
\end{align}

\begin{figure}[tbp]
    \centering
    \includegraphics[width=\linewidth]{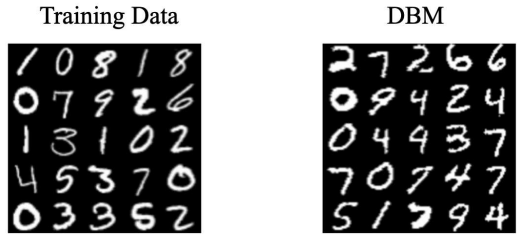}
    \caption{Left figure represents an example of the dataset of handwritten digits used to train DBM, and the right figure represents the data generated from DBM after training.}
    \label{fig:sample-generated}
\end{figure}

In our numerical experiments, we used the MNIST digit dataset of $6.0 \times 10^4$ handwritten images of 10 characters from $0$ to $9$ with $28 \times 28$ pixels, as shown in \dref{fig:sample-generated}. DBMs used for the experiments had a three-layer structure with $N_{0}=784$ visible units, $N_{1}=700$ hidden units in the first layer, and $N_{2}=700$ hidden units in the second layer. Following the conventional learning method \cite{salakhutdinov2009deep,salakhutdinov2010efficient}, we first pretrained DBM using the RBM-based layer-by-layer greedy learning method, and numerically maximized the log-likelihood by an approximate learning algorithm with a learning rate of the gradient descent method of $2.0 \times 10^{-3}$. The samples generated by DBM after training are shown in \dref{fig:sample-generated} and were confirmed to be close to the training dataset.

\begin{figure}[tbp]
    \centering
    \includegraphics[width=\linewidth]{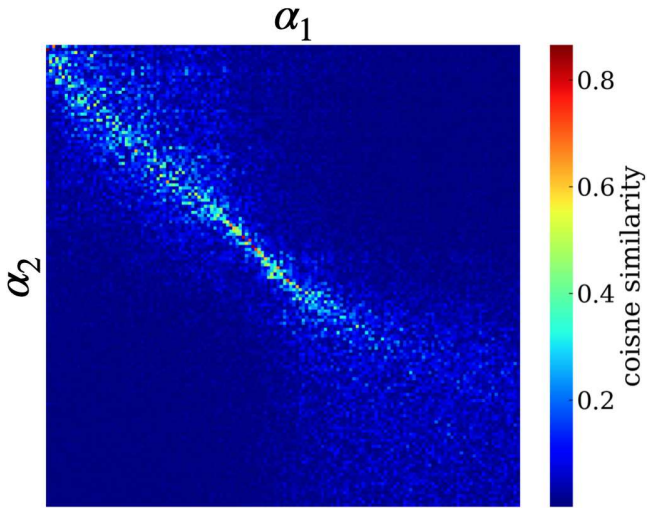}
    \caption{Heatmap of the interlayer correlations between the right singular vector $\{\B{v}^{(1)}_{\alpha_{1}}\}_{\alpha_{1}=1}^{700}$ of the learned weight parameter $\B{W}^{(1)}$ and the left singular vector $\{\B{u}^{(2)}_{\alpha_{2}}\}_{\alpha_{2}=1}^{700}$ of $\B{W}^{(2)}$.}
    \label{fig:inter-layer-corr}
\end{figure}

We performed the singular-value decomposition of the weight parameters $\B{W}^{(1)}$ and $\B{W}^{(2)}$ after the training, and first studied the cosine similarity between the right singular vectors $\{\B{v}_{\alpha_{1}}^{(1)}\}_{\alpha_{1}=1}^{700}$ of $\B{W}^{(1)}$ and left singular vectors $\{\B{u}_{\alpha_{2}}^{(1)}\}_{\alpha_{2}=1}^{700}$ $\B{W}^{(2)}$, referred to hereafter as ``interlayer correlation''. As shown in \dref{fig:inter-layer-corr}, interlayer correlations with large singular values are relatively large, whereas those with small singular values are almost zero. Furthermore, the interlayer correlation between modes with close indices, that is, $\alpha_{1} \approx \alpha_{2}$, is relatively large.

\begin{figure}[tbp]
    \centering
    \includegraphics[width=\linewidth]{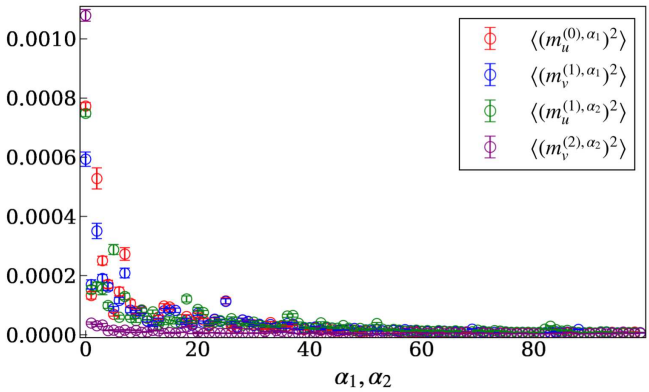}
    \caption{Singular-value index dependence of the second moments of overlaps between the singular vectors of weight parameters and the corresponding units for each layer.}
    \label{fig:dbm-average-units-svd-mode}
\end{figure}

\begin{figure}[tbp]
    \centering
    \includegraphics[width=\linewidth]{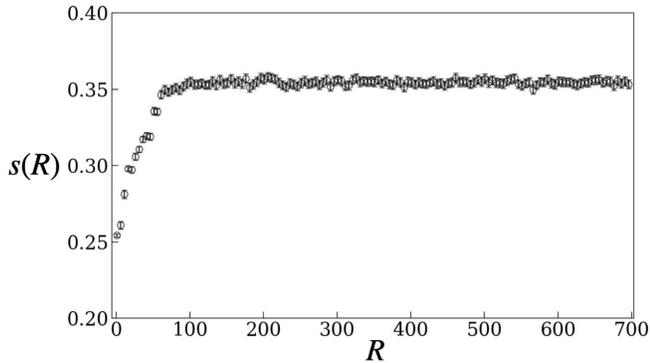}
    \caption{Averaged largest cosine similarity between train data and generated samples as a function of the rank number $R$ of the low-rank approximation.}
    \label{fig:sample-overlap-each-mode}
\end{figure}

Next, we focused on the overlaps between the singular vectors of each weight parameter and the corresponding units, defined by
\begin{equation*}
\begin{split}
    &m_{u}^{(0), \alpha_{1}} = \frac{1}{\sqrt{N_{0} N_{1}}} (\B{\sigma}^{(0)})^{\top} \B{u}_{\alpha_{1}}^{(1)},\\
    &m_{v}^{(1), \alpha_{1}} = \frac{1}{\sqrt{N_{0} N_{1}}} (\B{\sigma}^{(1)})^{\top} \B{v}_{\alpha_{1}}^{(1)},\\
    &m_{u}^{(1), \alpha_{2}} = \frac{1}{\sqrt{N_{1} N_{2}}} (\B{\sigma}^{(1)})^{\top}\B{u}_{\alpha_{2}}^{(2)},\\
    &m_{v}^{(2), \alpha_{2}} = \frac{1}{\sqrt{N_{1} N_{2}}} (\B{\sigma}^{(2)})^{\top} \B{v}_{\alpha_{2}}^{(2)}.
\end{split}
\end{equation*}
We then calculated the second-order moments of the four different overlaps with respect to the joint probability distribution of DBM after training. As shown in \dref{fig:dbm-average-units-svd-mode}, the overlaps with large singular values take finite values irrespective of the type of overlap, whereas the overlaps with small singular values are almost zero. In contrast, DBMs that do not function well as generators either have overlaps that are almost zero or only with the largest singular value finite. Therefore, for DBMs to function well as generators, it is important that the overlaps of multiple singular vectors be simultaneously finite. This is similar to the case of RBMs suggested in previous studies\cite{decelle2018thermodynamics,decelle2017spectral}, where
the generator performance is dominated by a few singular modes with large singular values\cite{decelle2018thermodynamics}. Therefore, we examined whether the same property holds for DBMs through numerical experiments. For this purpose, numerical experiments were performed using the following procedure:
\begin{enumerate}
  \item Reconstruct weight parameters by singular-value decomposition of the learned weight parameters up to rank $R$.
  \item Generate samples from DBM with the weight parameter reconstructed as the rank $R$.
  \item For each sample, calculate the cosine similarity for each training data and record its largest value.
  \item Calculate the average of largest value over all samples and denote it as $s(R)$.
\end{enumerate}
Our numerical result for $s(R)$ is shown as a function of rank number $R$ in \dref{fig:sample-overlap-each-mode}. The rank number dependence of $s(R)$ is approximately expressed as
\begin{equation}
    s(R)-s(\infty) = A \exp \left(-\frac{R}{R^{\ast}}\right),
\end{equation}
where the characteristic rank $R^\ast$ is evaluated as $R^{\ast}\approx 14.25$.
This suggests that, even if the rank number of the weight parameters increases, the expressive capability of DBMs remains unchanged and saturated. In other words, as in the case of RBMs, only the first few modes of the weight parameters after training are considered important for DBMs to function well as generators.
The results of these numerical experiments are almost independent of the number of units in the hidden layers and the number of hidden layers, implying that a DBM that works well as a generator generally has the same property, independent of network size.

\section{Model of DBM after learning}
\label{sec:model}

Based on the numerical results provided in the previous section,
the weight parameters of DBMs to be analyzed in statistical mechanics are given in the form of singular-value decomposition:
\begin{align}
  \label{eq:svd-weight}
  \B{W}^{(l)} &= \sum_{\alpha_{l}=1}^{K_{l}} w^{(l)}_{\alpha_{l}} \B{u}^{(l)}_{ \alpha_{l}}(\B{v}^{(l)}_{ \alpha_{l}})^{\top}
      + \B{r}^{(l)},
\end{align}
where $\B{u}^{(l)}_{\alpha_{l}}$ and $\B{v}^{(l)}_{\alpha_{l}}$ are the left and right singular vectors corresponding to the singular value $w^{(l)}_{\alpha _{l}}$, respectively. The first term represents the singular modes that are important for DBM to function well as a generator, and the second term represents the term consisting of the remaining singular modes that are expected to behave as noise. We then assume that $K_{l} = \mac{O}(N_{l}^{0})$, considering the fact that the number of important singular modes remains almost the same, independent of the network size, as shown in the previous section.

When the weight parameters of a DBM that functions well as a generator are given a singular-value decomposition as in \eref{eq:svd-weight}, the singular values and singular vectors $\B{u}^{(l)}\equiv \{\B{u}_{\alpha_{l}}^{(l)}\}_{\alpha_{l}=1}^{K_{l}}$ and $\B{v}^{(l)} \equiv \{\B{v}_{\alpha_{l }}^{(l)}\}_{\alpha_{l}=1}^{K_{l}}$, and the noise term $\B{r}^{(l)}$ explicitly depend on the individual training data and learning process.
We focused on the typical properties of DBMs rather than on specific properties that depend on such individual problems, and studied them using a statistical-mechanical method for random spin systems. Therefore, we studied the typical properties of DBMs for ensembles of elements of singular vectors and noise terms as random variables.

The random variables are denoted by $\B{u} = \{\B{u}^{(l)}\}_{l=1}^{L}$, $\B{v} = \{\B{v}^{(l)}\}_{l=1}^{L}$, and $\B{r} = \{\B{r}^{(l)}\}_{l=1}^{L}$. Specifically, the elements of the singular vectors $\B{u}$ and $\B{v}$ are assumed to follow a probability distribution with a mean of zero, and
variance covariances of the following values for any $l=1, \ldots, L$ and any $\alpha_{l} = 1, \ldots, K_{l}$:
\begin{align}
  \mathbb{E}[(v_{j, \alpha_{l}}^{(l)})^{2}] &= \frac{1}{N_{l}},~~~\forall j = 1, \ldots, N_{l}, \label{v_variance} \\
  \mathbb{E}[(u_{i, \alpha_{l}}^{(l)})^{2}] &= \frac{1}{N_{l-1}},~~~\forall i = 1, \ldots, N_{l-1} \label{u_variance}, \\
  \mathbb{E}[v_{j,\alpha_{l}}^{(l)}, u_{j, \alpha_{l+1}}^{(l+1)}] &=
  \begin{cases}
  \frac{\rho}{N_{l}}, &\alpha_{l} = \alpha_{l+1} \\
  0, &\alpha_{l} \neq \alpha_{l+1},
  \end{cases},
  \forall j =1, \ldots, N_{l}, \label{inter-layer_correlation}
\end{align}
where $\mathbb{E}[\cdots]$ means an average with respect to the assumed probability distribution.
This means that only the right and left singular vectors with arbitrary $\alpha_{l} = \alpha_{l+1}$ have finite correlation $\rho$, which corresponds to the fact that the interlayer correlations are particularly large at $\alpha_{l} \approx \alpha_{l+1}$, as discussed in the previous section. This interlayer correlation is because of the coupling between hidden layers as a result of learning, which is a property that does not exist in RBM\cite{decelle2018thermodynamics} and appears only in DBM.

We also assume that the elements of the second term, $\B{r}^{(l)}$, followed a Gaussian distribution:
\begin{equation}
\label{eqn:r-dist}
    \B{r}^{(l)}\in \mab{R}^{N_{l-1}\times N_{l}}:~ r_{ij}^{(l)}\sim \mac{N}\left(0, \frac{\sigma_{l}^{2}}{M_{l}}\right),
\end{equation}
where $M_{l} \coloneqq \sqrt{N_{l-1} N_{l}}$.

\section{Replica Analysis of DBMs}
\label{sec:replica}
In this section, we discuss a statistical-mechanical approach to the DBM model presented in the previous section, in which the energy function has random variables such as the singular vectors $\B{u}$ and $\B{v}$ and noise $\B{r}$. Assuming that the free energy density of DBM has a self-averaging property under the ensemble of singular vectors and noise, the typical free energy density is evaluated by the replica method as follows:
\begin{equation}
  \label{DBM_replica_method}
  f = - \lim_{N \to \infty} \frac{1}{\beta N} \lim_{n \to \infty} \pdif{}{n} \ln \left[Z^{n}(\B{u}, \B{v}, \B{r}, \beta) \right]_{\B{u}, \B{v}, \B{r}},
\end{equation}
where $\left[\cdots\right]_{\B{u}, \B{v}, \B{r}}$ denotes the ensemble average with respect to random variables $\B{u}$, $\B{v}$, and $\B{r}$, and $N = \sum_{l=0}^{L}N_{l}$.

For convenience, we introduced some macroscopic order parameters as
\begin{align}
  m_{u,a}^{(l-1),\alpha_{l}} &= \frac{1}{\sqrt{M_{l}}} (\B{\sigma}_{a}^{(l-1)})^{\top} \B{u}_{\alpha_{l}}^{(l)},\\
  &\forall l = 1, \ldots, L,~ \alpha_{l} = 1, \ldots, K_{l}, \notag\\
  m_{v,a}^{(l),\alpha_{l}} &= \frac{1}{\sqrt{M_{l}}} (\B{\sigma}_{a}^{(l)})^{\top} \B{v}_{\alpha_{l}}^{(l)},\\
  &\forall l = 1, \ldots, L,~ \alpha_{l} = 1, \ldots, K_{l},\notag \\
  q_{ab}^{(l)} &= \frac{1}{N_{l}} (\B{\sigma}^{(l)}_{a})^{\top} \B{\sigma}_{b}^{(l)},~~\forall l = 0, \ldots, L.
\end{align}
The macroscopic quantities $m_{u,a}^{(l-1), \alpha_{l}}$ and $m_{v,a}^{(l),\alpha_{l}}$
represent the overlap between the visible and hidden units, and the left/right singular vector. The Edwards--Anderson (EA) order parameter $q_{ab}^{(l)}$ represents the overlap between the replicated units for each layer.
Hereafter, these order parameters are collectively denoted for the replica index as
\begin{align}
  \B{m}_{u}^{(l-1), \alpha_{l}} &\coloneqq \left(m_{u, a}^{(l-1), \alpha_{l}} \right)_{a=1}^{n},\\
  &\forall l = 1, \ldots, L,~ \alpha_{l} = 1, \ldots, K_{l}, \notag \\
  \B{m}_{v}^{(l), \alpha_{l}} &\coloneqq \left(m_{v, a}^{(l), \alpha_{l}} \right)_{a=1}^{n},\\
  &\forall l = 1, \ldots, L,~ \alpha_{l} = 1, \ldots, K_{l},\notag \\
  \B{Q}^{(l)} &\coloneqq \left(q_{ab}^{(l)} \right)_{a, b=1}^{n},~~\forall l = 0, \ldots, L.
\end{align}
Using these order parameters, the replicated partition function $\left[Z^{n}(\btheta) \right]_{\B{u}, \B{v}, \B{r}}$ for $\kappa_{l} \coloneqq N_{l}/N_{0}$ can be expressed as
\begin{multline}
    \label{eq:zn_expan1}
    \left[Z^{n}(\B{\theta}) \right]_{\B{u}, \B{v}, \B{r}} = \\
    \int d\B{\Theta} \exp \Bigg(N_{0} \Bigg[ \sum_{l=1}^{L} \sqrt{\kappa_{l-1} \kappa_{l}} \Bigg( \frac{(\beta\sigma_{l})^{2}}{2} \mathrm{tr} \left(\B{Q}^{(l-1)} \B{Q}^{(l)} \right)\\
    + \sum_{\alpha_{l}} \beta w^{(l)}_{\alpha_{l}} \B{m}_{u}^{(l-1), \alpha_{l}} \B{m}_{v}^{(l), \alpha_{l}} \\
    - \frac{1}{\sqrt{M_{l}}} \left(\B{m}_{u}^{(l-1), \alpha_{l}} \hat{\B{m}}_{u}^{(l-1), \alpha_{l}} + \B{m}_{v}^{(l), \alpha_{l}}\hat{\B{m}}_{v}^{(l), \alpha_{l}} \right) \Bigg)\\
    + \sum_{l=0}^{L} \kappa_{l} \left(-\frac{1}{2} \mathrm{tr} \left( \B{Q}^{(l)} \hat{\B{Q}}^{(l)} \right) + \Psi^{(l)}(\hat{\B{Q}}, \hat{\B{m}}_{u}^{(l), \alpha_{l+1}}, \hat{\B{m}}_{v}^{(l), \alpha_{l}}) \right) \Bigg]\Bigg),
\end{multline}
where
\begin{multline*}
d\B{\Theta} \coloneqq \prod_{l=1}^{L}\prod_{a, \alpha_{l}}\frac{\sqrt{M_{l}} dm_{a, u}^{(l-1), \alpha_{l}} d\hat{m}_{a, u}^{(l-1), \alpha_{l}}}{2\pi} \frac{\sqrt{M_{l}}dm_{a, v}^{(l), \alpha_{l}} d\hat{m}_{a, v}^{(l), \alpha_{l}}}{2 \pi} \\
\times \prod_{l=0}^{L} \prod_{a, b} \frac{N_{l} dq_{ab}^{(l)} d\hat{q}_{ab}^{(l)}}{4 \pi},
\end{multline*}
and
\begin{multline*}
  \Psi^{(l)}(\hat{\B{Q}}, \hat{\B{m}}_{u}^{(l), \alpha_{l+1}}, \hat{\B{m}}_{v}^{(l), \alpha_{l}}) \coloneqq
  \log \Bigg[\sum_{\B{\sigma}^{(l)}}\exp \Bigg( \frac{1}{2} (\B{\sigma}^{(l)})^{\top} \hat{Q}^{(l)} \B{\sigma}^{(l)} \\
  + \left(\sum_{\alpha_{l+1}} u_{\alpha_{l+1}}^{(l+1)} (\hat{\B{m}}_{u}^{(l), \alpha_{l+1}})^{\top} + \sum_{\alpha_{l}} v_{\alpha_{l}}^{(l)} (\hat{\B{m}}_{v}^{(l), \alpha_{l}})^{\top} \right) \B{\sigma}^{(l)} \Bigg]_{\B{v}^{(l)}, \B{u}^{(l+1)}},
\end{multline*}
where we define $\B{\sigma}^{(l)} \coloneqq (\sigma_{a}^{(l)})_{a=1}^{n}$ and $v_{\alpha_{0}}^{(0)} = u_{\alpha_{L+1}}^{(l+1)} =0$ for index binary conditions. For further analysis, assuming replica symmetry (RS) and considering the limit of $N \to \infty$, the partition function of \eref{eq:zn_expan1} is expressed as
\begin{multline*}
    \mathrm{extr}_{\B{\Theta}} \exp \Bigg( N_{0} n \Bigg[\sum_{l=1}^{L} \sqrt{\kappa_{l-1} \kappa_{l}} \Bigg(\frac{(\beta\sigma_{l})^{2}}{2}\left(1 - q^{(l)}q^{(l-1)}\right) \\
    + \sum_{\alpha_{l}} \beta w^{(l)}_{\alpha_{l}} m_{u}^{(l-1), \alpha_{l}} m_{v}^{(l), \alpha_{l}}\\ - \frac{1}{\sqrt{M_{l}}} \sum_{\alpha_{l}} \left(m_{u}^{(l-1), \alpha_{l}} \hat{m}_{u}^{(l-1), \alpha_{l}} + m_{v}^{(l), \alpha_{l}} \hat{m}_{v}^{(l), \alpha_{l}} \right)  \Bigg)  \\
    + \sum_{l=0}^{L} \kappa_{l} \Bigg(\frac{\hat{q}^{(l)}(q^{(l)}-1)}{2} \\
    + \left[\log 2 \cosh h_{l}(z, v^{(l)}_{\alpha_{l}}, u^{(l+1)}_{\alpha_{l+1}})\right]_{z, \B{v}^{(l)}, \B{u}^{(l+1)}}\Bigg)\Bigg]\Bigg),
\end{multline*}
with
\begin{equation}
  h_{l}(z, \B{v}^{(l)}, \B{u}^{(l+1)}) =
  \sqrt{\hat{q}^{(l)}} z + \sum_{\alpha_{l}} \hat{m}^{(l), \alpha_{l}}_{v} v^{(l)}_{ \alpha_{l}} + \sum_{\alpha_{l+1}} \hat{m}^{(l), \alpha_{l+1}}_{u} u^{(l+1)}_{\alpha_{l+1}},\nn
\end{equation}
where $\mathrm{extr}_{\B{\Theta}}$ means to take an extreme value with respect to $m_{u}^{(l), \alpha_{l+1}}$, $m_{v}^{(l), \alpha_{l}}$, $q^{(l)}$, $\hat{m}_{u}^{(l), \alpha_{l+1}}$, $\hat{m}_{v}^{(l),\alpha_{l}}$ and $\hat{q}^{(l)}$, and $[\cdots]_{z}$ means the average over the random variable $z \sim \mathcal{N}(0, 1)$.
The extreme-value conditions on the conjugate variables yield the following relations:
\begin{align}
  \hat{m}^{(l-1) \alpha_{l}}_{u} &= \beta \sqrt{M_{l}} w^{(l)}_{\alpha_{l}} m^{(l), \alpha_{l}}_{v}, \nn\\
  \hat{m}^{(l),\alpha_{l}}_{v} &= \beta \sqrt{M_{l}} w^{(l)}_{\alpha_{l}} m^{(l-1) \alpha_{l}}_{u} , \nn \\
  \hat{q}^{(l)} &= \beta^{2} \left(\sqrt{\frac{\kappa_{l-1}}{\kappa_{l}}} q^{(l-1)} \sigma_{l}^{2} + \sqrt{\frac{\kappa_{l+1}}{\kappa_{l}}} q^{(l+1)} \sigma_{l+1}^{2} \right). \nn
\end{align}
Furthermore, the extreme-value conditions of the order parameters lead to the following self-consistent equation:
\begin{align}
  m^{(l),\alpha_{l+1}}_{u} &= \left(\frac{\kappa_{l}}{\kappa_{l+1}} \right)^{\frac{1}{4}} \left[u^{(l+1)}_{\alpha_{l+1}} \tanh h_{l}(z, \B{v}^{(l)}, u^{(l+1)}_{\alpha_{l+1}}) \right]_{z, \B{v}^{(l)}, \B{u}^{(l+1)}}, \label{eq:saddle-eq-mu} \\
  m^{(l), \alpha_{l}}_{v} &= \left(\frac{\kappa_{l}}{\kappa_{l-1}} \right)^{\frac{1}{4}} \left[ v^{(l)}_{\alpha_{l}} \tanh h_{l}(z, \B{v}^{(l)}, \B{u}^{(l+1)}) \right]_{z, \B{v}^{(l)}, \B{u}^{(l+1)}}, \label{eq:saddle-eq-mv} \\
  q^{(l)} &= \left[ \tanh^{2} h_{l}(z, \B{v}^{(l)},\B{u}^{(l+1)}) \right]_{z,\B{v}^{(l)},\B{u}^{(l+1)}}. \label{eq:saddle-eq-q}
\end{align}
where $h_{l}(z, \B{v}^{(l)}, \B{u}^{(l+1)})$ is explicitly given by
\begin{multline}
  h_{l}(z, \B{v}^{(l)}, \B{u}^{(l+1)}) \coloneqq \\ \beta \Bigg(\sqrt{\left(\frac{\kappa_{l-1}}{\kappa_{l}}\right)^{\frac{1}{2}} \sigma_{l}^{2} q^{(l-1)} + \left( \frac{\kappa_{l+1}}{\kappa_{l}}\right)^{\frac{1}{2}} \sigma_{l+1}^{2} q^{(l+1)}} z \\
  + \sqrt{M_{l}} \sum_{\alpha_{l}} w^{(l)}_{\alpha_{l}} m^{(l-1),\alpha_{l}}_{u} v^{(l)}_{\alpha_{l}} \\
  + \sqrt{M_{l+1}}\sum_{\alpha_{l+1}} w^{(l+1)}_{ \alpha_{l+1}}m_{v}^{(l+1),\alpha_{l+1}} u^{(l+1)}_{\alpha_{l+1}}\Bigg),
\end{multline}
and $[\cdots]_{z,\B{v}^{(l)},\B{u}^{(l+1)}}$ is the expected value with respect to the Gaussian random variable $z$ and singular vectors. These self-consistent equations are reduced to those of RBMs\cite{decelle2018thermodynamics} when only $\kappa_{l}$ and $\kappa_{l+1}$ are finite and the others are zero.

\section{Phase Diagram of 3-layer DBMs}
\label{sec:PD}
In this section, we derive the phase diagram of 3-layer DBMs, which is the simplest model with a hierarchical structure under the RS assumption.

\subsection{Definition of Phases}
\label{sec:def_phases}
First, we define the phases characterized by the order parameters mentioned above. The model has five phases depending on the values of the order parameters, as follows:
\begin{itemize}
    \item a paramagnetic phase(P) :
    \begin{align}
      m_{u}^{(l-1), \alpha_{l+1}} &= m_{v}^{(l), \alpha_{l}} = 0,\\
      &\forall l=1, \ldots, L,~\forall \alpha_{l} = 1,~ \ldots, K_{l}, \notag \\,
      q^{(l)} &= 0,~~\forall l=0, \ldots, L.\nn
    \end{align}
    \item a spin glass phase(SG) :
    \begin{multline}
    m_{u}^{(l-1), \alpha_{l}} = m_{v}^{(l), \alpha_{l}} = 0, ~~ \forall l=1, \ldots, L,~\forall \alpha_{l} = 1,~ \ldots, K_{l}, \\,
    q^{(l)} \neq 0,~~\exists l=0, \dots, L. \nn
  \end{multline}
    \item a ferromagnetic phase(F) :
    \begin{multline}
    m_{u}^{(l-1),\alpha_{l}} \neq 0 \vee m_{v}^{(l), \alpha_{l}} \neq 0, \\ \exists l = 1, \ldots, L,~\exists \alpha_{l} = 1, \ldots, K_{l} \nn.
  \end{multline}
    \item a layer combination phase(LC)
    \begin{multline}
      m_{u}^{(l-1),\alpha_{l}}, m_{v}^{(l), \alpha_{l}} \neq 0 \nn,
      \\ \forall l = 1, \ldots, L,~~\exists \alpha_{l} = 1, \ldots, K_{l}
    \end{multline}
    \item a composition phase(CM) :
    \begin{multline}
        m_{u}^{(l-1), \alpha_{l}}, m_{v}^{(l), \alpha_{l}} \neq 0,
        \\ \forall l = 1, \ldots, L,~~\exists \alpha_{l} = 1, \ldots, K_{l}, \nn
    \end{multline}
    \begin{multline}
     m_{u}^{(l-1), \beta_{l}} \neq 0 \vee m_{v}^{(l), \beta_{l}} \neq 0 \nn,
     \\ \exists l = 1, \ldots, L, \exists \beta_{l} = 1, \ldots, K_{l} \setminus \{\alpha_{l}\}.
    \end{multline}
\end{itemize}
These phases can be interpreted in terms of generators as follows:
\begin{itemize}
  \item P: DBMs generate random samples;
  \item SG: DBMs generate samples that are not related to singular vectors of weight parameters;
  \item F: DBMs generate samples that are related to singular vectors of weight parameters;
  \item LC: DBMs generate samples that are related to singular vectors of the same singular mode;
  \item CM: DBMs generate samples that are related to singular vectors corresponding to at least two or more singular modes.
\end{itemize}

Numerical experiments have suggested that for DBMs to function well as generators, the overlaps should be simultaneously finite. Therefore, the ferromagnetic phase, the layer combination phase and the composition phase are necessary conditions for multiple singular vectors to have a finite overlap simultaneously.
Note, however, that since the ferromagnetic phase includes the layer combination and the composition phase, the ferromagnetic phase is a loose condition for reproducing the state as in the numerical experiments.

\subsection{Instability of paramagnetic phase}
From the linear stability analysis of Eq. \eqref{eq:saddle-eq-mu} to \eqref{eq:saddle-eq-q} around the paramagnetic solution, the transition temperature between the paramagnetic phase and spin glass phase $T_{\text{P-SG}}$ is obtained as
\begin{equation}
    \frac{T_{\text{P-SG}}}{\sqrt{\sigma_{1} \sigma_{2}}} = \left(\frac{\sigma^{2}_{1}}{\sigma^{2}_{2}} + \frac{\sigma^{2}_{2}}{\sigma^{2}_{1}} \right)^{\frac{1}{4}}.
\end{equation}
A similar instability analysis leads to the transition temperature between the paramagnetic and ferromagnetic phases $T_{\mathrm{P-F}}$ as follows:
\begin{multline}
          \frac{T_{\mathrm{P-F}}}{\sqrt{\sigma_{1} \sigma_{2}}} = \frac{1}{\sqrt{2 \sigma_{1} \sigma_{2}}} \Bigg((w^{(1)}_{1})^{2} + (w^{(2)}_{1})^{2} \\
          + \sqrt{\left((w^{(1)}_{1})^{2} - (w^{(2)}_{1})^{2} \right)^{2} + 4 \rho (w^{(1)}_{1})^{2} (w^{(2)}_{1})^{2} }\Bigg)^{\frac{1}{2}}.
\end{multline}
The SG-F phase boundary is also derived as
\begin{multline}
    \frac{T_{\text{SG-F}}}{\sqrt{\sigma_{1} \sigma_{2}}} = w_{1}^{(1)} \left(\frac{\left[\text{sech}^{2} h_{1}(z, 0, 0) \right]_{z}}{2 \sigma_{1} \sigma_{2}} \right)^{\frac{1}{2}} \\
    \times \Bigg( \left[\text{sech}^{2} h_{0}(z, 0, 0) \right]_{z} + \left(\frac{w^{(2)}_{1}}{w^{(1)}_{1}} \right)^{2} \left[\text{sech}^{2} h_{2}(z, 0, 0) \right]_{z} \\
    + \Bigg(\Bigg( \left[\text{sech}^{2} h_{0}(z, 0, 0) \right]_{z} - \Bigg(\frac{w^{(2)}_{1}}{w^{(1)}_{1}} \Bigg)^{2}\left[\text{sech}^{2} h_{2}(z, 0, 0) \right]_{z}\Bigg)^{2} \\
    + 4 \rho \Bigg(\frac{w^{(2)}_{1}}{w^{(1)}_{1}} \Bigg)^{2} \left[\text{sech}^{2} h_{0}(z, 0, 0) \right]_{z} \left[\text{sech}^{2} h_{2}(z, 0, 0) \right]_{z} \Bigg)^{\frac{1}{2}} \Bigg)^{\frac{1}{2}}.
\end{multline}
Note that these transition temperatures do not depend on the details of the distribution functions of singular vectors $\B{u}$ and $\B{v}$. Rather, as we shall see later, the specific shape and correlation of the distribution function affect the nature of the ferromagnetic phase, including the layer combination and composition phases.

\begin{figure}[tbh]
    \centering
    \includegraphics[width=80mm]{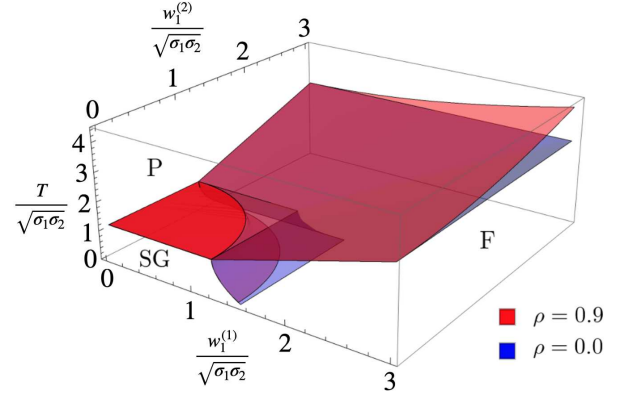}
    \caption{Phase diagram of $3$-layer DBMs when $\sigma_{1} = \sigma_{2} =1$ and $N_{0} = N_{1} = N_{2}$ with two different values of the interlayer correlation $\rho$. The red surface represents the phase boundary when $\rho = 0.9$, and the blue surface represents the phase boundary when $\rho = 0.0$.}
    \label{fig:phase-diag-cor}
\end{figure}

First, \dref{fig:phase-diag-cor} shows a phase diagram with two different values of interlayer correlation $\rho$ when $\sigma_{1}=\sigma_{2}=1.0$ and $N_{0} = N_{1} = N_{2}$. As shown in \dref{fig:phase-diag-cor}, the ferromagnetic phase expands, and the spin-glass phase contracts as the interlayer correlation $\rho$ increases.

\begin{figure}[tbh]
    \centering
    \includegraphics[width=85mm]{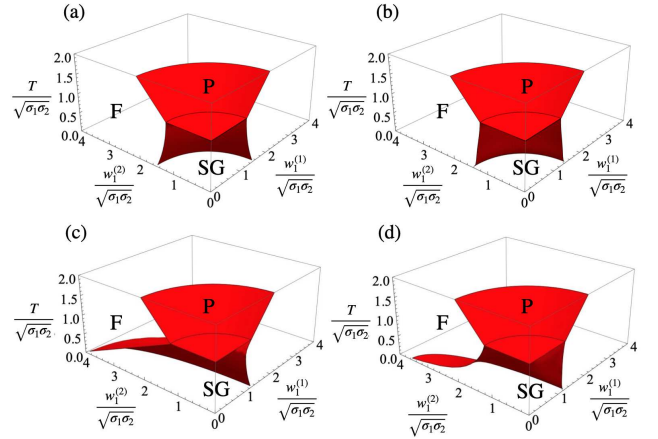}
    \caption{
Phase diagram of 3-layer DBM at $\sigma_{1} = \sigma_{2} = 1.0$ with 4 different number of units. The surface represents the phase boundary of SG-P, F-P and F-SG at $\kappa_{1} = \kappa_{2} = 1.0$(a), $\kappa_{1} = 1.0 \times 10^{-4}, ~\kappa_{2} = 1.0$(b), $\kappa_{1} =1.0,~ \kappa_{2} = 1.0 \times 10^{-4}$(c) and $\kappa_{1} = 1.0 \times 10^{-4},~ \kappa_{2} = 1.0 \times 10^{-4}$(d).}
    \label{fig:phase-diag-cor2}
\end{figure}

Next, we study the dependence of the phase diagram on the number of units in the hidden layers. \dref{fig:phase-diag-cor2} shows the phase diagram depending on the number of units in each layer.
As shown in the figure, the topology of the phase diagram is very sensitive to the ratio of the number of elements in each layer, that is, $\kappa_1$ and $\kappa_2$, and a re-entrant phase transition also appears at $\kappa_{1} = 1.0$, $\kappa_{2} = 1.0 \times 10^{-4}$, and $\kappa_{1} = 1.0 \times 10^{-4}$, $\kappa_{2} = 1.0 \times 10^{-4}$. This does not appear in RBM analysis\cite{decelle2018thermodynamics} and is a feature of DBMs.

The results show that the ferromagnetic phase expands and the spin-glass phase contracts as the number of middle layer units $N_{1}$ decreases, rather than when the number of last layer units $N_{2}$ decreases. It is suggested that, to function well as a generator, the number of units in the middle layer $N_{1}$ should be reduced rather than that of units in the final layer $N_{2}$.
Furthermore, when the number of units $N_{2}$ is reduced, the ferromagnetic phase becomes larger when $N_{1}$ is small. This result suggests that larger networks are not necessarily better and that it is important to determine the number of parameters appropriately for each layer.

\subsection{Properties of Layer Combination Phase}
\begin{figure}[h]
\centering
\includegraphics[width=80mm]{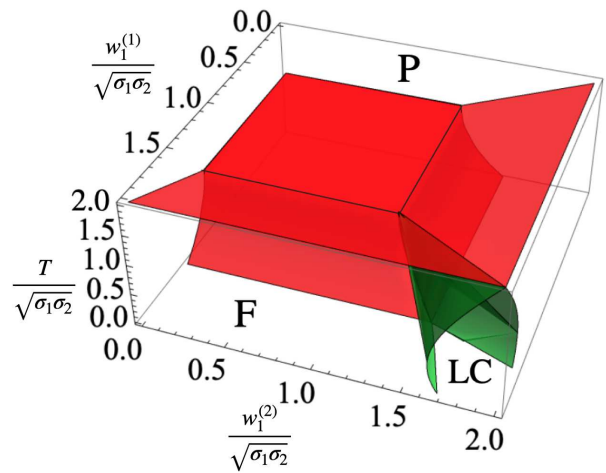}
\caption{
Phase diagram of 3-layer DBM without interlayer correlation at $\sigma_{1} = \sigma_{2} = 1.0$ and $\kappa_{1} = \kappa_{2} =1.0$.
The green surface represents the phase boundary between the ferromagnetic (F) phase and the layer combination (LC) phase.}
\label{fig:lc-phasediag}
\end{figure}

In this subsection, we discuss the properties of the layer combination phase, where overlaps with the largest singular modes of all layers emerge simultaneously.
We find that the layer combination phase is always the ferromagnetic phase when the interlayer correlation is finite, whereas when the interlayer correlation is zero, the stability conditions of the layer combination phase are determined separately from that of the ferromagnetic phase as
\begin{equation}
\left(\frac{T}{w^{(1)}_{1}}\right)^{2}\ge \left[\mathrm{sech}^{2} h_{0}(z, 0, 0)\right]_{z} \Big[ \text{sech}^{2} h_{1}(z, 0, u^{(2)}_{1}) \Big]_{z, u^{(2)}_{1}},
\end{equation}
and
\begin{equation}
\left(\frac{T}{w^{(2)}_{1}}\right)^{2} \ge \left[\mathrm{sech}^{2} h_{2}(z, 0, 0)\right]_{z} \left[\text{sech}^{2} h_{1}(z, v^{(1)}_{1}, 0) \right]_{z, v^{(1)}_{1}}.
\end{equation}
This means that the LC-F phase boundary depends on the specific shape of the distribution function of each largest singular vector. Assuming a multivariate Gaussian distribution that satisfies Eqs. \eqref{v_variance} to \eqref{inter-layer_correlation}, we obtain the phase diagram shown in \dref{fig:lc-phasediag}.
Even without interlayer correlation, the layer combination phase certainly exists in the phase diagram, but it is restricted to the region where ${w}^{(1)}_{1}$ and ${w}^{(2)}_{1}$ are close to each other.
This result implies that interlayer correlation is not essential for the existence of the layer combination phase, but the region is restricted.
It is suggested that finite interlayer correlations are important for the existence of a layer combination phase that effectively utilizes all the layers, or, in the absence of the interlayer correlation, that the largest singular values of each layer are close to each other.

\subsection{Properties of Composition Phase}
Our numerical experiments on DBM, shown in \dref{fig:dbm-average-units-svd-mode}, indicate that when DBM works well as a generator and
the overlaps of the multiple singular vectors are simultaneously finite.
This phase is distinguished from the ferromagnetic phase and is defined as the composition phase in subsection~\ref{sec:def_phases}. Therefore,
the conditions for the appearance of the composition phase, which are more stringent than those for the ferromagnetic phase, are studied here.

Similarly, from the linear stability analysis of Eqs. \eqref{eq:saddle-eq-mu} to \eqref{eq:saddle-eq-q} around the state that only the singular vector corresponding to the largest singular value has finite overlap, the transition temperature between the ferromagnetic phase and composition phase $T_{\mathrm{CM-F}}$ is obtained as
\begin{multline}
  \label{eq:cm-con}
    \frac{T_{\text{CM-F}}}{\sqrt{\sigma_{1} \sigma_{2}}} =
    w_{2}^{(1)} \left(\frac{\left[\text{sech}^{2} h_{1}(z, v^{(1)}_{1}, u^{(2)}_{1}) \right]_{z, v^{(1)}_{1}, u^{(2)}_{1}}}{2 \sigma_{1} \sigma_{2}} \right)^{\frac{1}{2}} \\
    \times \Bigg( \left[\text{sech}^{2} h_{0}(z, 0, u^{(1)}_{1}) \right]_{z, u^{(1)}_{1}} \\
    + \left(\frac{w^{(2)}_{2}}{w^{(1)}_{2}} \right)^{2} \left[\text{sech}^{2} h_{2}(z, v^{(2)}_{1}, 0) \right]_{z, v^{(2)}_{1}} \\
    + \Bigg(\Bigg( \left[\text{sech}^{2} h_{0}(z, 0, u^{(1)}_{1}) \right]_{z, u^{(1)}_{1}} \\
    - \Bigg(\frac{w^{(2)}_{2}}{w^{(1)}_{2}} \Bigg)^{2}\left[\text{sech}^{2} h_{2}(z, v^{(2)}_{1}, 0) \right]_{z, v^{(2)}_{1}}\Bigg)^{2} \\
    + 4 \rho \Bigg(\frac{w^{(2)}_{2}}{w^{(1)}_{2}} \Bigg)^{2} \left[\text{sech}^{2} h_{0}(z, 0, u^{(1)}_{1}) \right]_{z, u^{(1)}_{1}} \\
    \times \left[\text{sech}^{2} h_{2}(z, v^{(2)}_{1}, 0) \right]_{z, v^{(2)}_{1}} \Bigg)^{\frac{1}{2}} \Bigg)^{\frac{1}{2}}.
\end{multline}
Note that the CM-F phase boundary also depends on the specific shape of the distribution function of each singular vector with the largest singular value. For this purpose, we assumed a specific functional form of the probability distribution $p(\B{u}, \B{v})$. First, we draw the phase diagram by assuming a multivariate Gaussian distribution with the properties of Eqs. \eqref{v_variance} to \eqref{inter-layer_correlation}, as in the previous section. As a result, the composition phase does not appear and the state in which only finite overlaps of singular vectors corresponding to the largest singular value always dominates, which is similar to the previous study on RBMs \cite{decelle2018thermodynamics}.

\begin{figure}[tbh]
    \centering
    \includegraphics[width=80mm]{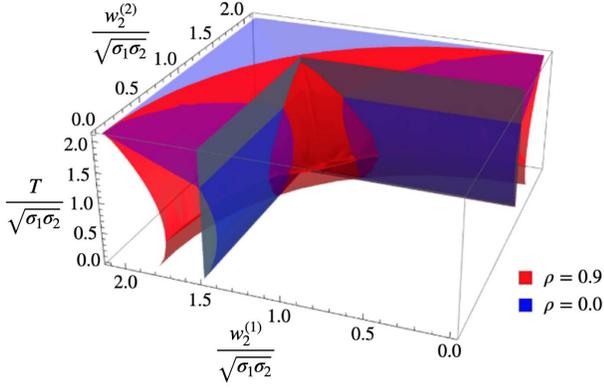}
    \caption{
Phase diagram of 3-layer DBM with two different values of the interlayer correlation $\rho$ when $\sigma_{1} = \sigma_{2} =1.0$, $\kappa_{1} = \kappa_{2} = 1.0$ and $w^{(1)}_{1}/\sqrt{\sigma_{1} \sigma_{2}} = w^{(2)}_{1}/\sqrt{\sigma_{1} \sigma_{2}}=1.5$.
The black surface represents the region where the largest singular value and the second singular value are interchanged, and the phase boundary expressed in \eref{eq:cm-con} has no meaning outside the black surface.
The blue and red surfaces represent the CM-F phase boundary at $\rho=0.0$ and $\rho=0.9$, respectively.
 The composition phase exists in the region bounded by the black and red or blue surfaces.
}
    \label{fig:cm-phase-diag}
\end{figure}

Assuming a multivariate Laplace distribution instead of the Gaussinan distribution with the same properties, the phase diagram can be evaluated to show the existence of the composition phase appeared as seen in \dref{fig:cm-phase-diag}.
Then, it is important that the largest singular value and the second singular value are close to each other for the composition phase to appear. Furthermore, in the presence of interlayer correlations, the composition phase is restricted to the region where $w^{(1)}_{2}/\sqrt{\sigma_{1} \sigma_{2}}$ and $w^{(2)}_{2}/\sqrt{\sigma_{1} \sigma_{2}}$ have almost the same value, although it appears in the region where $(w^{(1)}_{2}/\sqrt{\sigma_{1} \sigma_{2}})^{2} + (w^{(2)}_{2}/\sqrt{\sigma_{1} \sigma_{2}})^{2}$ is small compared with the case where the interlayer correlations are small. This means that although the region where the composition phase exists is limited, the composition phase appears even when the largest singular value and the second singular value are not close to each other, compared with the case where the interlayer correlation is relatively small. Furthermore, the composition phase expanded with an increase in the largest singular value.

\section{Phase Diagram of General $L$ layers DBM}
\label{sec:infinite}
In the previous section, focusing on the 3-layer DBMs, we have discussed the dependence on the number of units and interlayer correlations.
In the following section, we study the dependence of each phase on the number of layers by varying the number of layers. For the analysis, we first assume that the singular value $w^{(l)}_{\alpha_{l}}$ of the weight parameter $W^{(l)}$ and variance $\sigma_{l} $ of the noise term $\B{r}^{(l)}$ satisfy the following conditions:
\begin{align}
    w^{(1)}_{\alpha_{1}} = w^{(2)}_{\alpha_{2}} = &\cdots = w^{(L-1)}_{\alpha_{L-1}} \equiv w_{\alpha}. \\
    \sigma_{1} = \sigma_{2} = &\cdots = \sigma_{L-1} \equiv \sigma.
\end{align}
Subsequently, the self-consistent equations in Eqs. \eqref{eq:saddle-eq-mu} to \eqref{eq:saddle-eq-q} are given by
\begin{align}
  m^{(l),\alpha}_{u} &= \left(\frac{\kappa_{l}}{\kappa_{l+1}} \right)^{\frac{1}{4}} \left[u^{(l+1)}_{ \alpha} \tanh h_{l}(z, \B{v}^{(l)}, \B{u}^{(l+1)}) \right]_{z, \B{v}^{(l)}, \B{u}^{(l+1)}} \label{eq:saddle_eq_LDBM_mu}, \\
  m^{(l), \alpha}_{v} &= \left(\frac{\kappa_{l}}{\kappa_{l-1}} \right)^{\frac{1}{4}} \left[v^{(l)}_{\alpha} \tanh h_{l}(z, \B{v}^{(l)}, \B{u}^{(l+1)}) \right]_{z, \B{v}^{(l)}, \B{u}^{(l+1)}} \label{eq:saddle_eq_LDBM_mv}, \\
  q^{(l)} &= \left[ \tanh^{2} h_{l}(z, \B{v}^{(l)}, \B{u}^{(l+1)}) \right]_{z, \B{v}^{(l)}, \B{u}^{(l+1)}} \label{eq:saddle_eq_LDBM_q},
\end{align}
where $h_{l}$ is defined as
\begin{multline}
  h_{l}(z, \B{v}^{(l)}, \B{u}^{(l+1)}) \coloneqq \\ \beta \Bigg(\sqrt{\left(\frac{\kappa_{l-1}}{\kappa_{l}}\right)^{\frac{1}{2}} q^{(l-1)} + \left( \frac{\kappa_{l+1}}{\kappa_{l}}\right)^{\frac{1}{2}} q^{(l+1)}} \sigma z \\
  + \sqrt{M_{l}} \sum_{\alpha} w_{\alpha} m^{(l-1),\alpha}_{u} v^{(l)}_{\alpha}
  + \sqrt{M_{l+1}}\sum_{\alpha} w_{\alpha}m_{v}^{(l+1),\alpha} u^{(l+1)}_{\alpha} \Bigg).
\end{multline}
In the following, we evaluate the phase diagram of the $L$-layer DBMs based on these self-consistent equations.
\subsection{Dependence of phase diagram on the number of layers}
\begin{figure}[h]
    \centering
    \includegraphics[width=85mm]{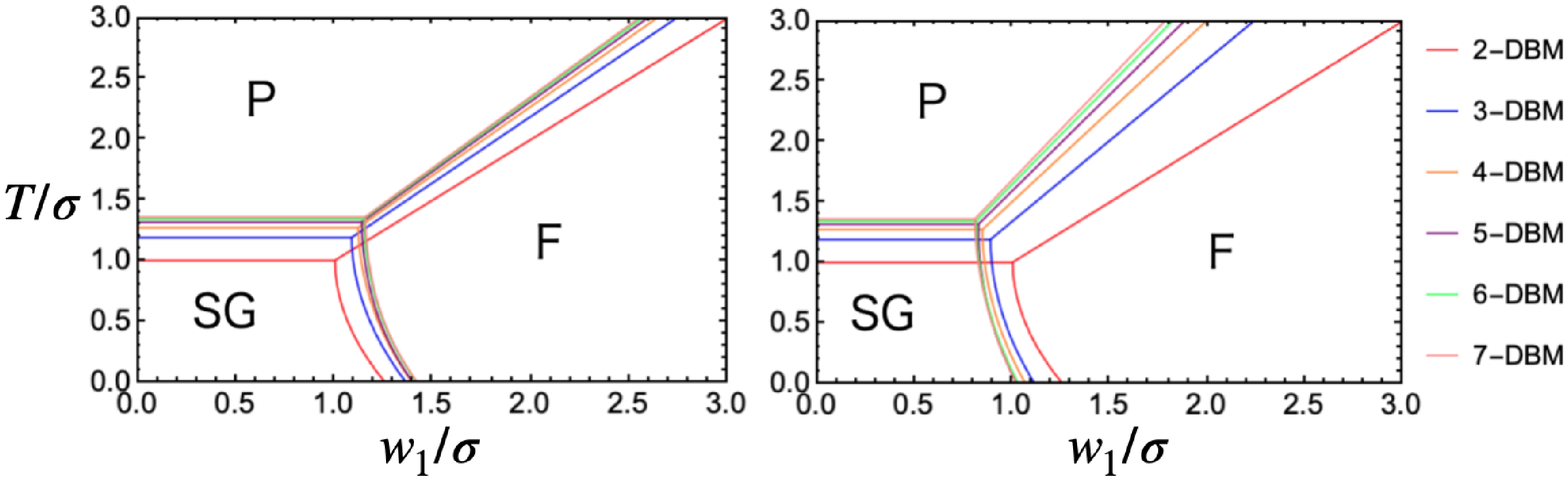}
    \caption{
Phase diagram of $L$-layer DBM in the plane of temperature and the largest singular value for several $L$ with the interlayer correlation $\rho=0.2$ (left) and $0.8$(right).
}
    \label{fig:L-layers-DBM-phase-diagram}
\end{figure}

As in the analysis of the $3$-layer DBM,
the P-SG phase boundary for the $L$-layer is derived
from the linear stability analysis of Eqs. \eqref{eq:saddle_eq_LDBM_mu} to \eqref{eq:saddle_eq_LDBM_q}, which is reduced to the condition that the largest eigenvalue of the tridiagonal matrix $G^L_{\rm{P-SG}}$ equals 1, where
\begin{equation}
\label{eq:general-L-P-SG-Matrix}
G^L_{\rm{P-SG}}=
\left(\frac{\sigma}{T}\right)^{2}
    \begin{pmatrix}
    0 & \left(A_{0}^{1} \right)^{2}& \cdots& 0 \\
    \left(A_{1}^{0} \right)^{2} & \ddots & \ddots & \\
    \vdots      & \ddots &        & \left(A_{L-2}^{L-1} \right)^{2}\\
    0 &   & \left(A_{L-1}^{L-2} \right)^{2} & 0
    \end{pmatrix},
\end{equation}
with $A_{l}^{l^{\prime}} \equiv \left(\frac{\kappa_{l^{\prime}}}{\kappa_{l}}\right)^{\frac{1}{4}}$.
When $\kappa_{0} = \kappa_{1} = \cdots = \kappa_{L-1}$ as a special case, i.e., when the number of units in all layers is uniform, the largest eigenvalue can be obtained explicitly.
Consequently, the P-SG transition temperature for $L$-layer DBM is evaluated as
\begin{equation}
\label{eq:L-layer-P-SG-Line}
    \frac{T^{L}_{\text{P-SG}}}{\sigma} = \sqrt{2 \cos \left(\frac{1}{L+1} \pi \right)}.
\end{equation}
The P-F phase boundary is also analyzed from the linear stability analysis and determined by the condition that the largest eigenvalue of the pentadiagonal matrix $G^L_{\rm{P-F}}$ equals $1$, where
\begin{align}
  \label{eq:general-L-P-F-Matrix}
  &G^L_{\rm{P-F}} = \notag \\
  &\frac{w_{1}}{T}\begin{pmatrix}
      0  & 1 & 0  &  & &   \\
      1  & 0  & 0 & \rho A_{0}^{2}   &  & \\
      \rho A_{2}^{0}  & 0 & 0 & 1 &  &   \\
      & \ddots &  \ddots & \ddots    & \ddots &  \\
      & & 1 & 0 & 0 &  \rho A_{L-3}^{L-1}\\
      & &  \rho A_{L-1}^{L-3} & 0 &0 & 1  \\
      0 &   &  &0 &1 &  0
    \end{pmatrix}.
\end{align}
Even with $\kappa_{0} = \kappa_{1} = \cdots = \kappa_{L-1}$, the eigenvalues of \eref{eq:general-L-P-F-Matrix} cannot be obtained; however, the P-F phase boundary is derived within a first-order perturbation of $\rho$ to an approximation with
\begin{equation}
\label{eq:L-layer-P-F-Line}
    \frac{T^{L}_{\text{P-F}}}{\sigma} \approx \frac{w_{1}}{\sigma} \left(1 + \rho \cos \left(\frac{\pi}{L}  \right)  \right).
\end{equation}
The SG-F boundary is also determined by the condition that the largest eigenvalue of the pentadiagonal matrix $G^L_{\rm{SG-F}}$ is $1$, where
\begin{align}
  \label{eq:general-L-F-SG-Matrix}
  &G^L_{\rm{SG-F}} = \notag \\
      &\frac{w_{1}}{T} \begin{pmatrix}
      0  & \gamma_{1} & 0  &  & &   \\
       \gamma_{2}  & 0  & 0 & \rho \gamma_{2} A_{0}^{2}  &  & \\
      \rho \gamma_{2} A_{2}^{0} & 0 & 0 & \gamma_{2} &  &   \\
       & \ddots &  \ddots & \ddots    & \ddots &  \\
       & & \gamma_{L-1} & 0 & 0 & \gamma_{L-1} \rho A_{L-3}^{L-1} \\
       & & \gamma_{L-1} \rho A_{L-1}^{L-3} &0 &0 &  \gamma_{L-1}  \\
      0 &   &  &0 & \gamma_{L} &  0
      \end{pmatrix}
\end{align}
and $\gamma_{l}$ is defined for $l=1,\ldots L$ as
\begin{equation}
    \gamma_{l}  \equiv \left[\mathrm{sech}^{2} h_{l}(z, 0, 0) \right]_{z}.  \nn
\end{equation}
The eigenvalue of \eref{eq:general-L-F-SG-Matrix} must be obtained numerically because $\gamma_{l}$ depends on $q^{(l)}$. The phase diagram is evaluated numerically for several layer numbers for two typical values of the interlayer correlation after learning in \dref{fig:L-layers-DBM-phase-diagram}. When the number of layers is increased, the SG-F phase boundary exhibits qualitatively different behavior depending on the value of the interlayer correlations. When the interlayer correlation is relatively small, the phase boundary shifts in the direction where $w_{1}/\sigma$ became small, and the ferromagnetic phase shrunk, whereas the ferromagnetic phase expands when the interlayer correlation is large. The threshold of interlayer correlation depends on the number of layers.

\subsection{infinite layer limit}
\begin{figure}[]
    \centering
    \includegraphics[width=85mm]{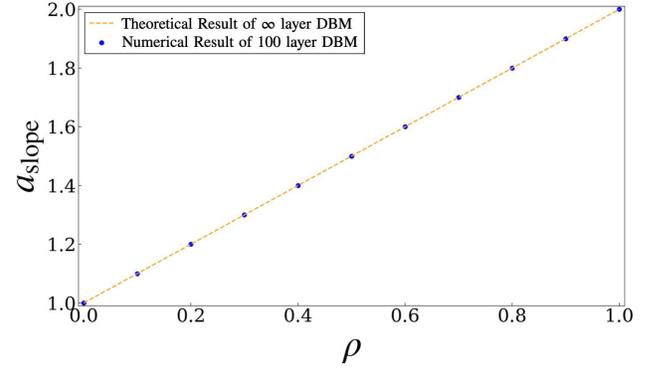}
    \caption{Slope $a_{\text{slope}}$ of the phase boundary between ferromagnetic and paramagnetic phases of $L$-layer DBM in the plane of temperature and the first singular value as a function of the interlayer correlation $\rho$. The dotted line is the result obtained by the first-order perturbation calculation of $\rho$ for $L=\infty$, and each point represents numerical result for $L=100$ layer DBM.}
    \label{fig:numerical-F-M-line2}
\end{figure}

In this subsection, we study the phase diagram of $L=\infty$-DBMs.
By taking the limit of $L \to \infty$ in \eref{eq:L-layer-P-SG-Line} and \eref{eq:L-layer-P-F-Line}, the P-SG and P-F phase boundaries are obtained as
\begin{equation}
    \label{eq:inf-layer-P-SG-Line}
     \frac{T^{\infty}_{\text{P-SG}}}{\sigma} = \sqrt{2},
\end{equation}
and
\begin{equation}
    \label{eq:inf-layer-P-F-Line}
     \frac{T_{\text{P-F}}^{\infty}}{\sigma} \approx \left(1 + \rho \right) \frac{w_{1}}{\sigma},
\end{equation}
respectively.
Although \eref{eq:inf-layer-P-F-Line} is justified only for small $\rho$,
the numerical results shown in \dref{fig:numerical-F-M-line2} suggest that \eref{eq:inf-layer-P-F-Line} holds when $L$ is large, even if the condition that $\rho$ is sufficiently small is not satisfied.
The phase boundary converges to a finite region within the limit of an infinite number of layers.

The critical point of the $\infty$-layer DBMs is determined from Eq. (\ref{eq:inf-layer-P-SG-Line}) and (\ref{eq:inf-layer-P-F-Line}) as follows:
\begin{equation}
  \label{eq:triple-point-inf-layer-DBM}
    \left(\frac{w_{1}}{\sigma}, \frac{T}{\sigma}  \right) = \left(\frac{\sqrt{2}}{\rho + 1}, \sqrt{2}  \right),
\end{equation}
while that of the RBM, the 2-layer DBM, is given by
\begin{equation}
\label{eq:triple-point-RBM}
    \left(\frac{w_{1}}{\sigma}, \frac{T}{\sigma}  \right) = \left(1, 1  \right).
\end{equation}
Therefore, the tricritical point for the interlayer correlation $\rho^{\ast}=\sqrt{2}-1$ increases perpendicular to the temperature and the first singular-value plane from the point of 2-layer DBMs to the point of the $\infty$-layer. Numerical results suggest that this vertical variation of the tricritical point is also valid for other finite-layer DBMs. Thus, the behavior of the tricritical point changes at the boundary of $\rho^{\ast}$, which gives the threshold of the interlayer correlation at the boundary between the expanding and shrinking ferromagnetic phases as $L$ increases, as discussed in the previous subsection.
More specifically, when the interlayer correlation after learning is smaller than $\rho^{\ast}$, the deeper networks of the DBM lead to the contraction of the ferromagnetic phase.

This result implies that naively increasing the number of layers does not necessarily play a positive role from the perspective of the DBM functioning well as a generator.
Even if the interlayer correlation can be learned to be larger than the threshold $\rho^{\ast}$, the phase boundary will converge as the number of layers increases; thus, it is necessary to carefully consider whether it is beneficial to naively increase the number of layers from the viewpoint of necessary conditions.

\section{Conclusion}
\label{sec:summary}
In this work, based on the results of numerical experiments on DBMs, the typical properties of DBMs in more realistic settings are analyzed using  a statistical mechanical framework to clarify the characteristics that make a DBM function well as a generators.
In our model setting, the ferromagnetic phase can be interpreted as a phase in which the generator functions well because of its finite overlap with the singular vectors. Therefore, the parameter domain of the ferromagnetic phase is regarded as a condition for DBMs to function well.
This framework describes the typical properties of DBMs, such as the phase diagram,  by means of the self-consistent equations governing the behavior of the order parameters.
Our analysis reveals, as one of the unversal results of DBMs, that the phase boundary of the ferromagnetic phase depends only on the second moments of the probability distribution of the element of the given singular vector and not on the details of the distribution more than the second moments.
We remark below on what can be learned from the present study about the number of units in each layer and the layer number in the DBM configuration, respectively.

Applying the framework to a 3-layer DBMs, in which a hierarchical structure appears for the first time, it was found that increasing the number of units in each layer does not necessarily enlarge the ferromagnetic phase. This suggests that  it is important to consider which layer to increase the number of units in, rather than simply increasing the overall number of units.
Our study also indicated that for the DBMs to function well as a generator, the singular values of the weight parameters in each layer must be close values to each other.

We also found that the interlayer correlation of the weight parameters, appearing only in the hierarchical structure, plays an important role in determining the properties of DBMs.
When the interlayer correlation is smaller than a threshold value, the ferromagnetic phase shrinks as the number of layers increases, while when the correlation is larger than the threshold value, the ferromagnetic phase expands. This means that the performance of DBMs does not necessarily improve with increasing the number of layers when the interlayer correlation is smaller than the threshold value.
This threshold value is explicitly obtained by analyzing DBMs with an infinite number of layers in our setting. During the learning process, it may be difficult to control the interlayer correlations of the weight parameters, but it is possible to observe them from time to time. From the above viewpoint, it would be meaningful to observe the interlayer correlation during the learning process, using our estimated threshold value as a guide for comparison.

In this study, typical and universal properties of DBMs that are independent of the characteristics of the training data, were investigated using the statistical mechanical approach. Such a study based on typical evaluation takes advantage of the statistical mechanical approach. However, it is not possible to address how the weight parameters of DBMs are acquired because they are given as a probabilistic model at the beginning of the study.
A possible future study would be %Another natural research direction is
to investigate the typical properties of learning dynamics from the viewpoint of statistical mechanics, similar to the previous studies on RBMs\cite{decelle_spectral, decelle2018thermodynamics}. %In particular, investigating the impact of interlayer-correlations, which are newly emerged due to the hierarchical structure, and their dynamics. And
Some of the interesting issues, then, are to identify the effect of interlayer correlations on the learning dynamics, of which this study has revealed the importance, and the nature of the learning dynamics  near the threshold.

\begin{acknowledgments}
This work was supported by MEXT as “Program for Promoting Researches on the Supercomputer Fugaku” (DPMSD, Project ID: JPMXP1020200307). One of the authors, YI, was supported by WINGS-FMSP program at the Graduate School of Mathematical Science, the University of Tokyo.
\end{acknowledgments}

\appendix

\bibliographystyle{jpsj}

\end{document}